\documentclass[oneside, a4paper, onecolumn, 11pt]{article}
\usepackage{multirow}
\usepackage[left=1.5cm,top=2cm,bottom=2cm,right=1.5cm]{geometry}
\usepackage[utf8]{inputenc}
\usepackage{graphicx} 		% Add graphics capabilities
\usepackage{amsmath}  		% Better maths support
\usepackage{natbib}	% bibliography style
\usepackage{eurosym}
\usepackage{longtable}
\usepackage[font=small,labelfont=bf]{caption}

\usepackage{multirow}
\usepackage[left=1.5cm,top=2cm,bottom=2cm,right=1.5cm]{geometry}
\usepackage[utf8]{inputenc}
\usepackage{graphicx} 		% Add graphics capabilities
\usepackage{amsmath}  		% Better maths support
\usepackage{natbib}	% bibliography style
\usepackage{eurosym}
\usepackage{longtable}
\usepackage[font=small,labelfont=bf]{caption}

\begin{document}
\begin{center}
\LARGE{\textbf{Magnetic control of metafluids for fluid-like applications of metamaterials}
}
\end{center}
%\vspace{5 pt}
%\title{}
\vspace{3 pt}
\begin{center}
\large{Ezra Ben-Abu$^1$,
       Anna Zigelman$^1$,
       Sefi Givli$^1$, and
       Amir D. Gat$^{1,*}$}\\
\vspace{5 pt}
% \address{%
\small{$^1$Faculty of Mechanical Engineering, Technion - Israel Institute of Technology, Haifa 3200003, Israel\\
$^*$ezrab@campus.technion.ac.il}
\end{center}

%\vspace{5 pt}
%\articlenote{A growing and maneuvering multi-stable material is based on the sequential activation of bi-stable elements via viscous flow.}
%\corremail{amirgat@technion.ac.il}
\vspace{3 pt}
\begin{abstract}
Metamaterials are structures composed of repeating unit-cells which enable macro-scale properties not found in nature. Since metamaterials are typically solid structures with predetermined interconnections, it is challenging to leverage their unique properties for many critical applications that require fluid-like behavior, such as heat engines or cooling cycles. Recent research suggested overcoming this limitation by creating a '\textit{mechanical metafluid}', which is a lubricated suspension of multistable unit-cells. However, realization of this concept necessitates the ability to control both velocity and state of the metafluid. Here, we propose the use of time-varying magnetic fields as a mechanism to manipulate metafluids. We focus on a lattice of magnetic multistable capsules enclosing gas and suspended within a liquid-filled tube. We derive the governing equations and examine one-dimensional fluid mechanics of the metafluid, both theoretically and experimentally, at the viscous limit under magnetic actuation. By applying time-varying magnetic fields, we control both the local compression and expansion of the capsules, as well as the entire flow field. Our theoretical results are compared with experimental data, showing good agreement. This work paves the way for the utilization of mechanical metamaterials to applications that require fluid-like behavior, thus extending the scope of metamaterial applications. 
\end{abstract}

% \keywords{growing materials; multistable structures; single-input actuation; viscous actuation}

% \maketitle
\section{Introduction}\label{sec1}
Metamaterials are artificial structures consisting of periodically arranged unit cells that deform, rotate, buckle, fold, or snap in response to mechanical loads. Typically, these unit cells are arranged in fixed positions within a lattice structure (see e.g., ~\cite{Kadic_2013,Bertoldi_2017} and many more), so that 
 the micro-level architecture enables metamaterials to provide exceptional macro-behavior, properties, and  functionalities that surpass those of standard materials~\cite{Bertoldi_2017}. Although the metamaterial concept was proposed for the first time by Veselago in 1968, who suggested that materials with negative values of magnetic and electric response could theoretically exist~\cite{KUMAR20223016}, it was only in the past two decades that metamaterials have started to gain significant attention, particularly for their ability to manipulate optical~\cite{Arpin_2010,Soukoulis_2011}, acoustic~\cite{Cummer_2016}, and thermal~\cite{Schittny_2013} fields as well as stress waves~\cite{Peng_2011,Lu_2009}, and
 that have highly unusual properties, such as a ``negative mass''~\cite{Correa_2015,Correa_2015b,Babaee_2013} or ultra-high stiffness to weight
 ratio~\cite{Zheng_2014}.

One of the most common categories of metamaterials is the field of \textit{multi-stable metamaterials}, which are utilized in a wide range of applications, such as robotic arms~\cite{zhang2023plug,kaufmann2022harnessing,wu2021stretchable}, growing structures~\cite{abu2023growing}, deployable structures~\cite{meeussen2023multistable,veksler2024fluid,melancon2021multistable,melancon2022inflatable}, and many more. Multistable metamaterials leverage four main types of multi-stable unit elements: curved beams~\cite{matia2015dynamics,pan2022novel,qiu2004curved,matia2017leveraging}, origami~\cite{miura19852,miura1993concepts,part1guest1994folding,part2guest1994folding,part3guest1996folding}, conical frusta~\cite{bende2018overcurvature,Breitman_2022,wo2022bending,abu2024hft,pan20193d,ilssar2022dynamics,ben2024reprogrammable}, and thin elastic shells~\cite{Bertoldi_2024,gorissen2020inflatable,van2023nonlinear,de2025electropneumatic}. Recent works~\cite{Peretz_2022,Peretz_2023,Bertoldi_2024} developed a new class of \textit{mechanical metafluids}, consisting of multi-stable unit-cells suspended in a fluidic medium, and thus have the ability to flow and adapt to the shape of their container without the need for a precise arrangement of their constituent elements. Such \textit{mechanical metafluids} are a natural candidate for use of the unique properties of metamaterials for critical applications that require fluid-like behavior, such as heat engines or cooling cycles.

Although the concept of non-mechanical metafluids has been previously explored in various domains, including optical~\cite{Yang_2016,Urzhumov}, electromagnetic~\cite{Sheikholeslami}, and wave manipulation~\cite{Liu_2020}, each offering distinct advantages, liquefying mechanical metamaterials into metafluids was only recently demonstrated.  Mechanical multistable metafluids were first introduced by Peretz et al.~\cite{Peretz_2022,Peretz_2023}, who suggested that multi-stable metafluids can be utilized for energy harvesting and storage, as well as for refrigeration cycles, by programming the density of the metafluid under atmospheric conditions. Due to the analogy between the snapping of metafluids to different states and the phase change in standard fluids, they were able to implement the Brayton cooling cycle, with an improvement in the coefficient of performance (COP) of the Brayton cooling cycle by 61\%. Later on, Djellouli et al~\cite{Bertoldi_2024} proposed to  realize a  ‘metafluid’ with programmable compressibility, optical behaviour and viscosity, by mixing deformable spherical capsules into an incompressible fluid. It was claimed that the resulting metafluid would facilitate the development of smart robotics systems, highly tunable logic gates, and optical elements with switchable characteristics. Realizing the concepts in ~\cite{Peretz_2022,Peretz_2023} and~\cite{Bertoldi_2024}, however, requires the ability to control both the velocity and the state of the metafluid. Currently, to the best of our knowledge, there is no practical method for externally actuating, moving, compressing, and expanding multiple stable unit cells within liquid.

In order to address this challenge and pave the way for the use of metamaterials in applications that require fluid-like motion, we propose the use of time-variable magnetic fields to actuate metafluids. By using external time-varying magnetic fields, we aim to control the movement, compression, and expansion of multi-stable capsules composing the metafluid, thereby controlling its local properties. Below we present both experimental and theoretical results for the motion of the metafluid under various frequencies and numbers of capsules. Using this approach, \textit{mechanical metafluids} can be leveraged for many applications that involve motion, compression, and expansion of their unit cells, such as heat pumps, refrigeration, energy harvesting and energy storage.

\section{Results}
\begin{figure}
\includegraphics[width=0.8\linewidth]{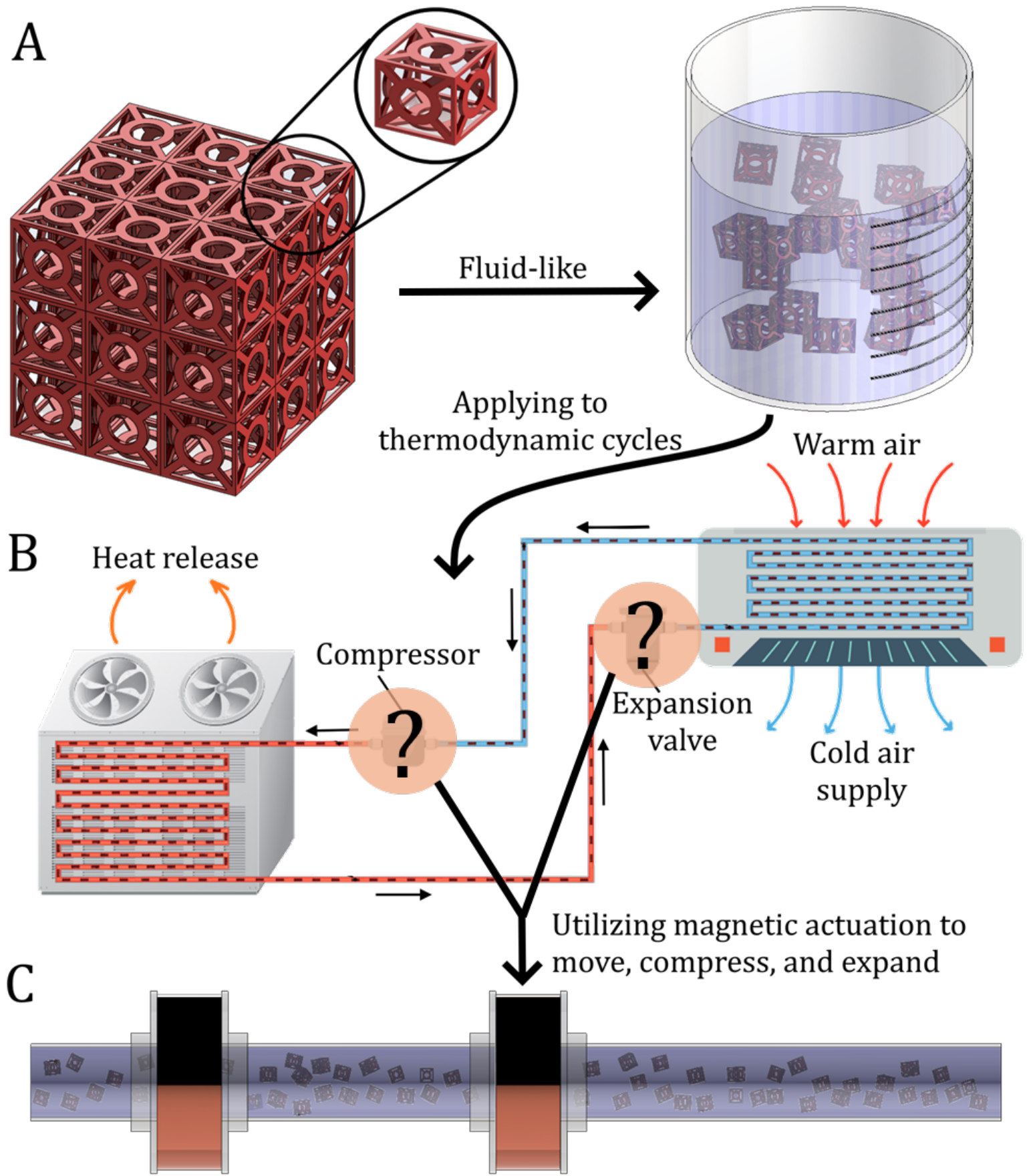}
\centering
\caption{{\bf{Liquifying metamaterial for a fluid-like applications.}} (A)  Metamaterial in a standard lattice structure (left) and metamaterial unit cells submerged in a liquid (right). (B) An example of an application of metafluid in a thermodynamic cycle, where motion, compression, and expansion of the unit cells are needed. (C) Our proposed concept for metafluid based on time-varying magnetic fields to manipulate the unit cells including motion, compression and expansion.}
\label{Fig1}
\end{figure}

Metamaterials are known for their exceptional properties that are not found in nature. However, many crucial applications still cannot benefit from these remarkable characteristics, as they require fluid-like behavior. Recent works demonstrated that it is possible to achieve fluid-like behavior of metamaterials by creating a suspension of multistable unit-cells ~\cite{Peretz_2022,Peretz_2023,Bertoldi_2024} (see Fig.~\ref{Fig1}(A)). However, applications involving multi-stable metafluid, which comprises multiple unit cells, require the ability to program its motion, configuration, and local mechanical properties, see for example Fig.~\ref{Fig1}(B). This cannot be achieved using traditional actuation methods, such as pumps and expansion valves, and thus necessitates external actuation. To address this limitation, we propose using time-varying magnetic fields to control the metafluids, as shown in Fig.~\ref{Fig1}(C).

Our proposed concept for the experimental setup (see Fig.~\ref{Fig2}(A)) consists of a long closed tube filled with metafluid, where six solenoids are installed at predetermined locations along the tube and are powered by separate power supplies and controlled by the drivers (for additional details regarding the drivers' connections, see Fig. 4 in Section 4 in SI). Moreover, the metafluid within the tube is created by suspending a 1D array of multistable elastic capsules in water. The capsules are filled with air under standard atmospheric conditions. Furthermore, both ends of each capsule are attached to the magnets and closed with a hemispherical cap to reduce friction. For additional details see the Experimental Section.

\subsection{Analysis}\label{S:motion}
\begin{figure}
\includegraphics[width=1\linewidth]{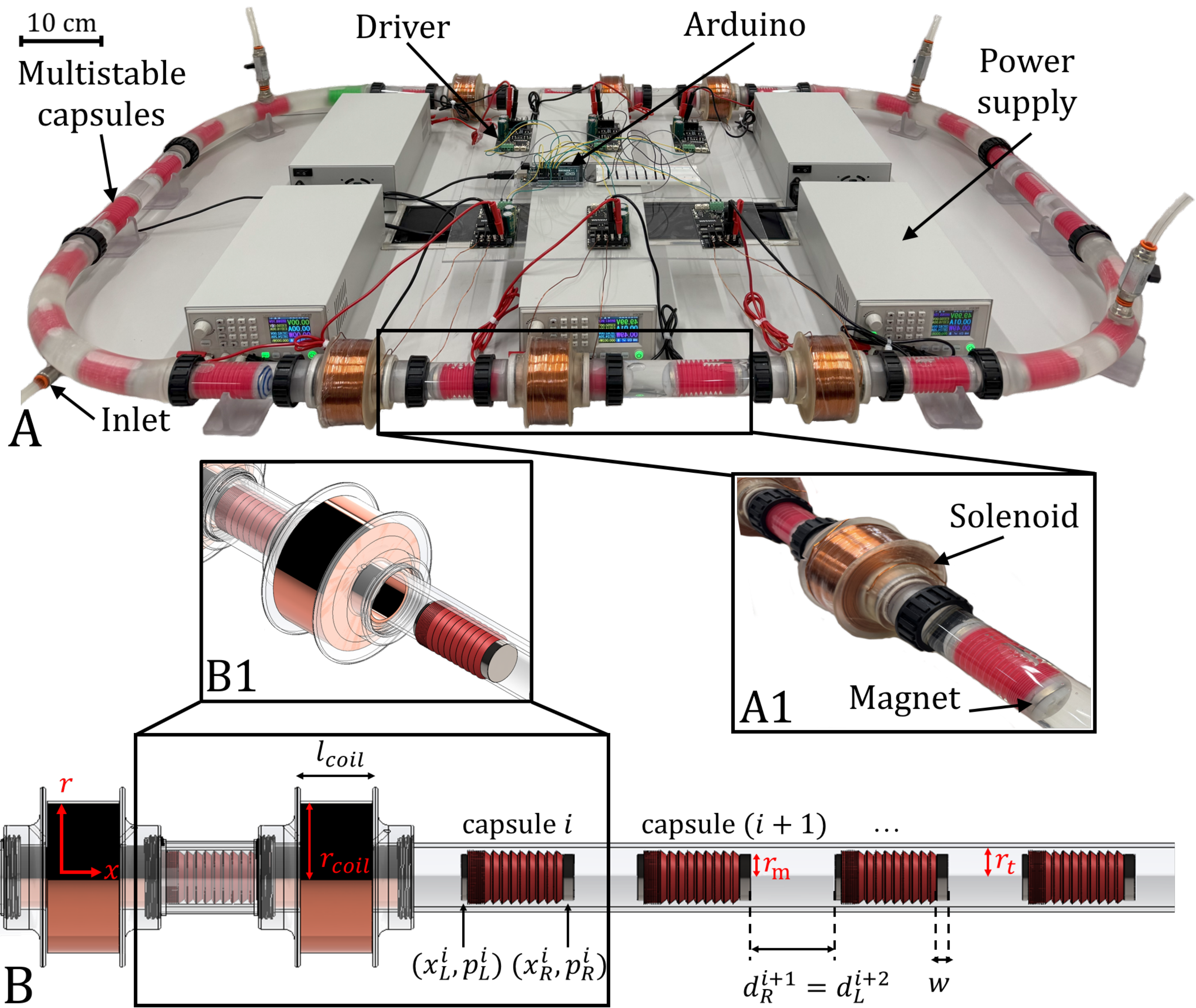}
\centering
\caption{{\bf{System configuration.}} (A) The image of the experimental setup, which comprises from a long closed tube filled with metafluid. The metafluid is obtained by suspending a 1D array of multistable elastic capsules in water. The capsules are equipped with magnets, which are attached to the capsules' ends and actuated by a magnetic field induced by six solenoids. The solenoids are installed over the tube at predefined locations, each connected to a power supply. The power supplies are connected to the drivers and all of the drivers are connected to Arduino, which gets signals from a computer according to our code (for more information see Section 4). In panel (A1), we show an increased view of the tube in the vicinity of the solenoid. (B) The sketch of our system, where our notations are indicated.  In panel (B1), we show an increased view of our system (rotated). The scale bar corresponds to 10 [cm].}
\label{Fig2}
\end{figure}

We adopt a Cartesian system of coordinates, with $x-$ axis passing along the center-line of the tube of length $L$. Let us assume that a capsule passes through the tube filled with fluid and which contains $J$ solenoids, whose centers are located at positions $s_j\in(0,L)$, $j=1,2,\ldots,J$. See Fig.~\ref{Fig2}(B).

%\begin{figure}
%\includegraphics[width=1\linewidth]{Sketch_Ezra.pdf}
%\centering
%\caption{The sketch of the system, where the physical parameters are indicated.}
%\label{Fig_4}
%\end{figure}

\begin{figure}
\includegraphics[width=1\linewidth]{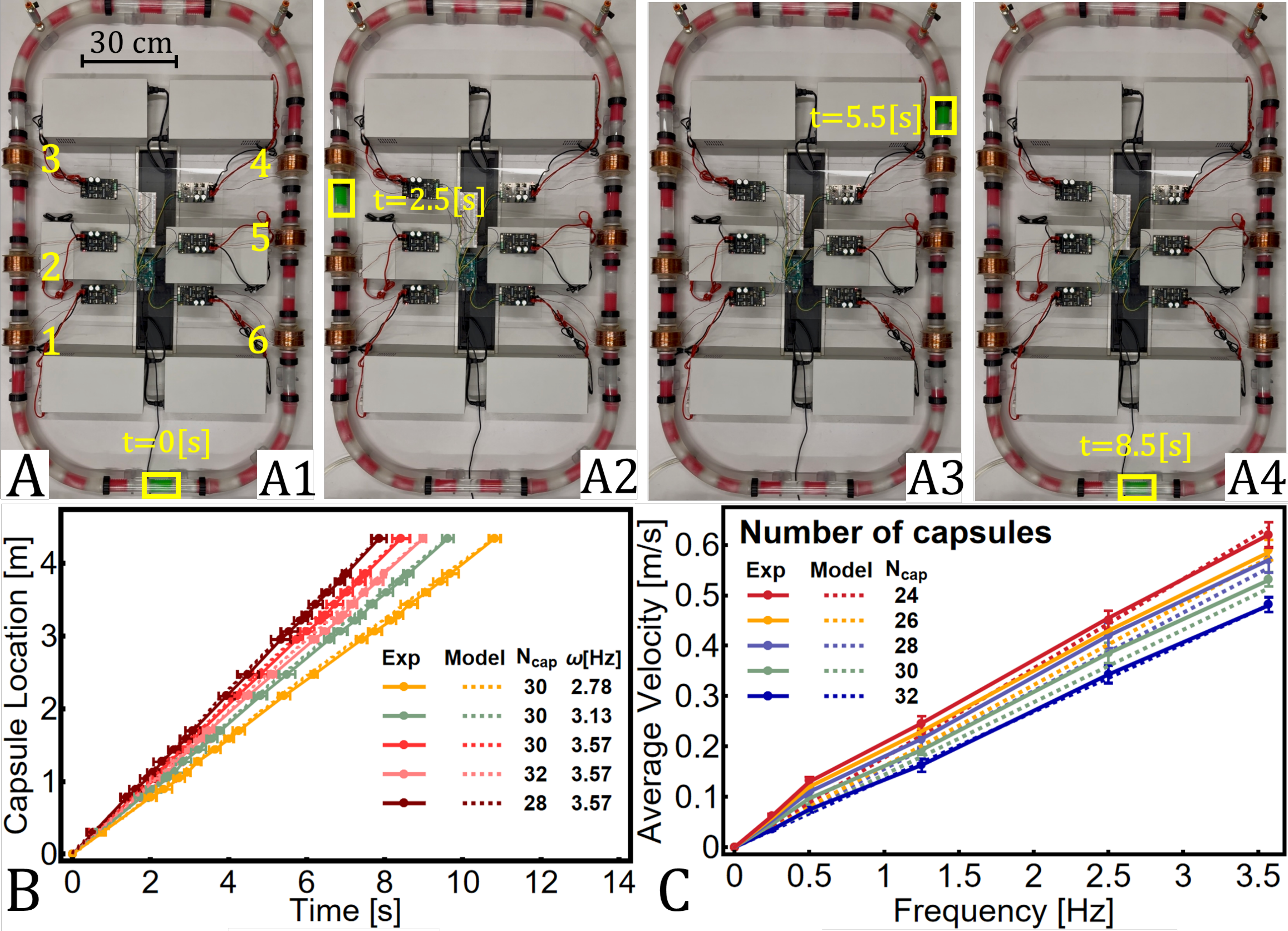}
\centering
\caption{{\bf{Controlling the metafluid velocity.}} 
(A) A sequence of frames from a video (see Movie S1) in our experiment, where we track the location of the green capsule at times $t=0,\,2.5,\,5.5$, and 8.5 seconds. Panels A1 and A4 look the same, which visualizes the fact that the capsules completed one cycle. The operation of the solenoids is with stepwise function with delay between the solenoids, as detailed in Experimental Section, where the current of the solenoids is $I_{\text{curr}}=16$ A, the number of capsules is $N_{\text{cap}}=30$, and the frequency of the actuated solenoids is $\omega=3.57$ Hz. The scale bar in panel (A1) corrsponds to 30 [cm]. (B)-(C) A comparison between theory (dashed lines) and experiment (solid lines) for (B) the location of capsules versus time for different frequencies and various numbers of capsules within the tube and for (C)  the mean velocity of the capsules versus frequency, for different numbers of the capsules within the tube. 
%Both (A) and (B) were obtained according to one cycle.
}
\label{Fig3}
\end{figure}

\begin{figure}
\includegraphics[width=1\linewidth]{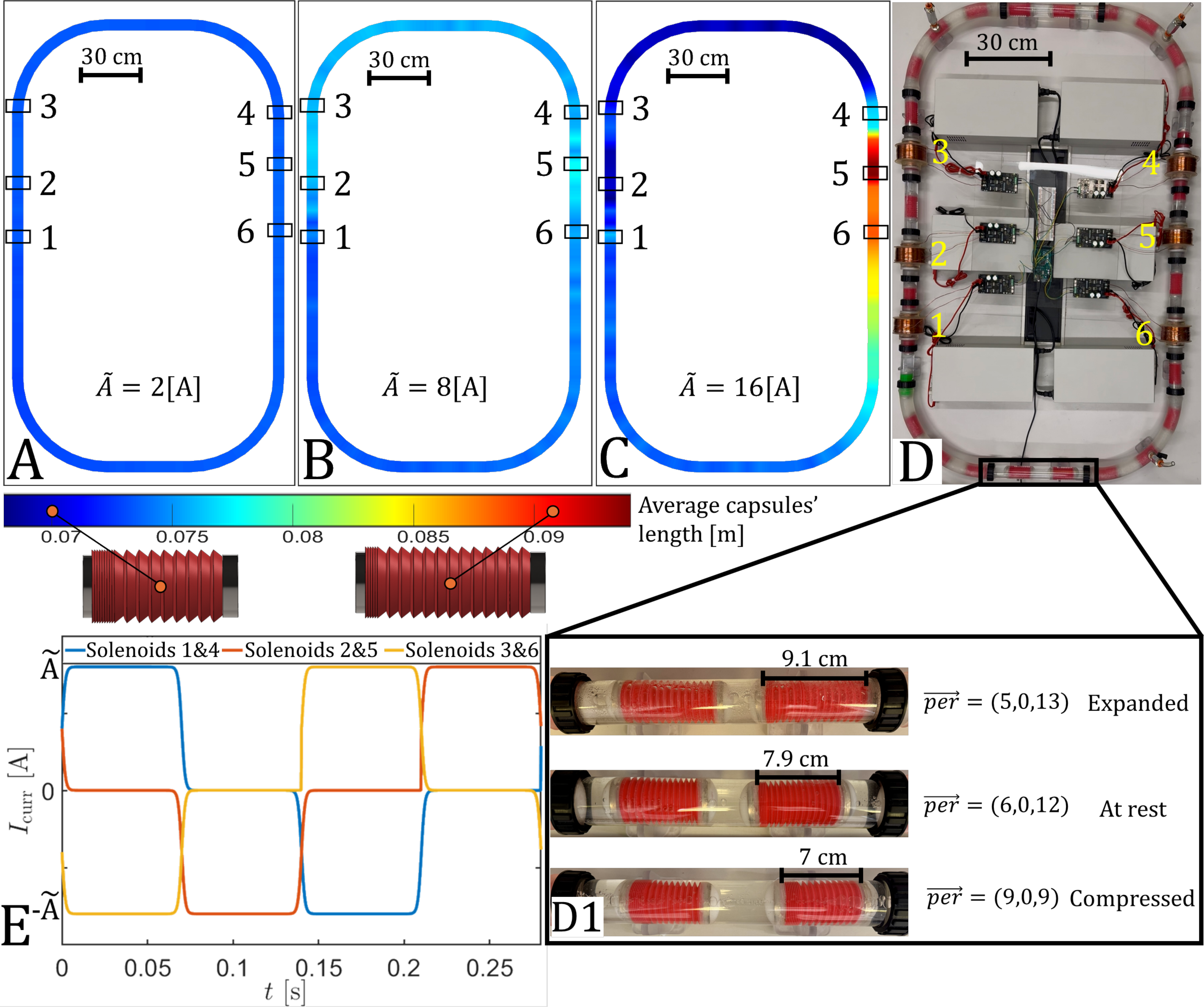}
\centering
\caption{{\bf{Controlling compression and expansion.}} (A)-(C) Average length map of the capsules, where initially they are assumed to be in an expanded state at permutation of $\overrightarrow{per}=(7,0,11)$, as a function of their location within the tube obtained by simulation, in three different cases of the current intensity, (A) $I_{\text{curr}}=2$ [A], (B) $I_{\text{curr}}=8$ [A], and (C) $I_{\text{curr}}=16$ [A]. 
The solenoids are marked with black rectangles and their numbering is indicated. (D) A frame from an experiment with the current of $I_{\text{curr}}=16$ [A] and frequency of $\omega=3.57$ Hz, with capsules' compression starting from the expanded state of the capsules. (D1) An increased view of the region marked by a black rectangle in (D), in three different states captured at different times: expanded, at rest, and compressed. (E) The actuation functions of the solenoids 1-3 with a general amplitude of $\tilde{A}$, which changes between the panels (A)-(C) as indicated. The solenoids 4-6 are actuated similarly to solenoids 1-3, in panel (C) and with opposite signs in panels (A)-(B). Both, in theory (panels (A)-(C)) and experiment, there 30 capsules in the system. The colorbar refers to panels (A)-(C), where below the colorbar we show for example two sketches of expanded and compressed capsules. 
The scale bars correspond to 30 [cm] in panels (A)-(D), and to 9.1, 7.9, and 7 [cm] in panel (D1).}
%The insets show the actuation function of the solenoids numbers 1, 2, and 3, whereas the solenoids 4, 5, and 6, operate either exactly as 1, 2, and 3, respectively, or as with the inverse sign of 1, 2, and 3, respectively.
\label{Fig4}
\end{figure}
% \begin{figure}
% \includegraphics[width=1\linewidth]{Fig1temp.pdf}
% \centering
% \caption{Sketch.}
% \label{Fig1temp}
% \end{figure}
Following~\cite{Peretz_2023}, we first introduce the momentum balance on the left and right sides of the capsule, in which magnetic forces between magnets of the neighboring capsules and between the coils and the magnets were incorporated,
\begin{subequations}\label{E:Force_Balance}
\begin{equation}\label{E:Force_Balance A}
\begin{aligned}
 p_L(t)\pi r_p^2-f_c\left(l_{\text{cap}}(t),t\right)-f_{\mu,L}(t)-p_M(t)\pi (r_p^2-r_e^2) 
 +\sum_{j=1}^J f_{m_j}(x_L(t)+w/2-s_j,t)+f_{d_L}(d_L) =0,
 \end{aligned}
\end{equation}
\begin{equation}\label{E:Force_Balance B}
\begin{aligned}
 -p_R(t)\pi r_p^2+f_c\left(l_{\text{cap}}(t),t\right)-f_{\mu,R}(t)+p_M(t)\pi (r_p^2-r_e^2) 
 +\sum_{j=1}^J f_{m_j}(x_R(t)-w/2-s_j,t)-f_{d_R}(d_R)=0,
 \end{aligned}
\end{equation}
\end{subequations}
and a mass balance
\begin{equation}\label{E:diff Q_L Q_R}
q_L-q_R=2\pi r_t(r_t-r_e)\left(\frac{\partial x_R}{\partial t}-\frac{\partial x_L}{\partial t}\right),
\end{equation}
where $p_L$, $p_M$, and $p_R$ denote the pressures at the left end, at the middle, and at the right end of the capsule, respectively, $r_e$ is the maximal radius of the frustum, $d_L$ and $d_R$, are the distances between the magnets on the left and right sides and the neighboring (from the left and from the right) magnets, respectively. Moreover, $f_c(l_{\text{cap}}(t),t)$ appearing in~\eqref{E:Force_Balance} is the elastic force exerted by the capsule, which is given by the multistable relation as introduced in~\cite{Peretz_2022,Peretz_2023} and multiplied by the area of the capsule's bases, namely 
\begin{equation}\label{E:elast force}
    f_c(l_{\text{cap}}(t),t)=\pi r_e^2 \left[a(t)l_{\text{cap}}(t)+b(t)\right],
\end{equation}
where $l_{\text{cap}}(t)=x_R(t)-x_L(t)$ denotes the capsule's length and $a(t)$, $b(t)$ denote two coefficients of the piece-wise linear function, so that their values depend on the permutation. For additional details regarding the definition of $f_c$, see Section 3.1 in SI.
Moreover, $f_{\mu,L}$ (and $f_{\mu,R}$) the viscous resistance acting on the left (and on the right) plate, under the assumption of annular geometry, is given by
%\begin{subequations}\label{E:def F_mu_s}
\begin{equation}\label{E:def F_mu_s}
    f_{\mu,L}=\frac{2\pi r_p w \mu}{r_t-r_p}\left[ \frac{\partial x_L}{\partial t}+\frac{(r_t-r_p)^2}{2\mu w}\left(p_M(t)-p_L(t)\right)\right],
\end{equation}
% \begin{equation}
%     f_{\mu,R}=\frac{2\pi r_p w \mu}{r_t-r_p}\left[ \frac{\partial x_R}{\partial t}+\frac{(r_t-r_p)^2}{2\mu w}\left(p_R(t)-p_M(t)\right)\right],
% \end{equation}
%\end{subequations}
where $w$ is the width of the magnet. The mass flux $q_L$ is given by lubrication approximation as
\begin{equation}\label{E:Flux}
 q_L= 2\pi r_t \left(
 \frac{(r_t-r_p)^3}{12\mu}\frac{p_L(t)-p_M(t)}{w}-\frac{r_t-r_p}{2}\frac{\partial x_L}{\partial t}
 \right).
\end{equation}
For the expressions for $F_{\mu,R}$ and $q_R$, see SI, Section 2. Moreover, the expressions for the forces between two neighboring magnets or between a magnet and solenoid $j$, $j=1,\ldots,J$, as well as the operation regimes of the solenoids are found in SI, Section 1.
%\begin{equation}\label{E:Q_R}
%   q_R= 2\pi r_t \left(
%     \frac{(r_t-r_p)^3}{12\mu}\frac{p_M(t)-p_R(t)}{w}-\frac{r_t-r_p}{2}\frac{\partial x_R}{\partial t}
%     \right).
% \end{equation}

%\subsection{A 1D lattice of capsules}\label{S:lattice}
The closed tube contains a 1D-lattice of $N_{\text{cap}}$ capsules. In our experiment, the system is closed (not exposed to the surrounding air) and the capsules are cyclically moving along the tube in a closed loop. 
%However, in our model, for the sake of simplicity (without loss of generality), the tube is assumed to be straight and parallel to the $x-$axis, but in order to simulate this cyclic motion in our model, we assume periodic boundary conditions. The periodicity here means that the rightmost capsule in the tube affects the leftmost capsule, as if it were to its left, and the leftmost capsule in the tube affects the rightmost capsule, as if it were to its right. For further details see Section 2 and Fig. 2 in SI. Moreover, in the calculation of the force between the coils and the magnets, the cyclic motion of the capsules is taken into account in the appropriate manner. 
% For detailed numerical algorithm including the construction of mass matrix and forcing term vector, see Section 3.2 in SI.     
Initially, ay $t=0$, we track each capsule (and define a number for each capsule), where we start counting the capsules from $x=0$, moving toward $x=L$. % The numbers of the capsules remain the same during the whole simulation (without any adjustment when any of the capsules passes through $x=L$).
%Each capsule (number $i$) in the array is in a stable permutation of the form $\overrightarrow{\mathbf{\text{per}}}_i=(n_o,0,n_c)$, where $n_o$, $n_s=0$, and $n_c$ denote the number of frusta in open state (I), spinodal (s), and close state (II), respectively. 
The variables defining the position and pressure (at both ends) of the capsule $i$ are $x^{i}_L$, $x^i_R$, $p^i_L$, $p^i_R$, $i\in \{1,2,...,N_{\text{cap}}\}$, where $i=1$ denotes the leftmost (initially) capsule and $i=N_{\text{cap}}$ denotes the rightmost (initially) capsule in the tube. Thus, in our system of equations for $N_{\text{cap}}$ capsules in the tube there are $4N_{\text{cap}}$ unknowns. Hence, we need to formulate a system of $4N_{\text{cap}}$ equations. 
Note that the above force balance \eqref{E:Force_Balance} and mass conservation equations \eqref{E:diff Q_L Q_R} yield overall $2N_{\text{cap}}$ equations. For more details see Section 2 in SI.

To formulate the remaining $2N_{\text{cap}}$ equations, additional relations are obtained using mass and momentum balance in the regions between adjacent capsules~\cite{Peretz_2023}. Since the viscous resistance between the capsules is negligible compared to the viscous resistance of the ends (where the magnets are located), we obtain that
%$P^i_{L}=P^{i-1}_{R}$, for $i=2,\ldots I$, so that in index notation we have the following equations,
\begin{equation}\label{E:P inside index}
    p^i_{L}=p^{i-1}_{R}, \quad i=2,\ldots,N_{\text{cap}} \quad \text{and} \quad p^{1}_L=p^{N_{\text{cap}}}_{R}.
\end{equation}
% where we used the periodicity boundary condition to relate between the first and the last capsules.
% \begin{equation}\label{E:P at node 1}
%     p^{1}_L=p^{N_{\text{cap}}}_{R}.
% \end{equation}
Requiring mass conservation in the gaps between the capsules yields that  %additional $N_{\text{cap}}-1$ equations,  namely 
\begin{equation}\label{E:mass conserv between capsules}
\begin{aligned}
q^i_{L}-q^{i-1}_{R}=\pi r_t^2 \left(\frac{\partial x^{i-1}_{R}}{\partial t}-\frac{\partial x^i_{L}}{\partial t}\right), \quad i=2,\ldots,N_{\text{cap}} 
\quad  \text{and} \quad q^1_{L}-q^{N_{\text{cap}}}_{R}=\pi r_t^2 \left(\frac{\partial x^{N_{\text{cap}}}_{R}}{\partial t}-\frac{\partial x^1_{L}}{\partial t}\right),
\end{aligned}
\end{equation}
where in~\eqref{E:P inside index} and~\eqref{E:mass conserv between capsules}, we used the periodicity boundary condition to relate between the first and the last capsules. 
% on~\eqref{E:mass conserv between capsules}, we obtain the last equation,
% \begin{equation}\label{E:mass conserv near boundaries}
% q^1_{L}-q^{N_{\text{cap}}}_{R}=\pi r_t^2 \left(\frac{\partial x^{N_{\text{cap}}}_{R}}{\partial t}-\frac{\partial x^1_{L}}{\partial t}\right).
% \end{equation}
%For more details see Section ?? in SI.

This problem is rendered dimensionless and transitions between multistable states are computed by solving it iteratively each time. More details regarding the transformation to dimensionless variables, formulation of the initial conditions, our methodology of solution, and the construction of the mass matrix and the forcing term vector are given in SI, Sections 2 and 3. Moreover, to reproduce our results, use Code C1.

% Note that the equation in~\eqref{E:temp V gen} is coupled to the system of 4$N_{cap}$ equations discussed above. 
% % in Sections~\ref{S:motion}--\ref{S:lattice}. 
% This coupling is reflected through the piecewise constant coefficients $a(t)$ and $b(t)$ that appear in the definition of the elastic force (see~\eqref{E:elast force}), since they depend on the temperature of the encapsulated gas. 

% When capsules open/close the encapsulated gas cools down/heats up. To take into account temperature variations due to volume changes, we use the following equation,
% \begin{equation}\label{E:temp V gen}
% \begin{aligned}
%     C_v m_{\text{gas}} \frac{d\mathcal{T}(t)}{dt}&+\frac{Rm_{gas}\mathcal{T}(t)}{V(t)}\frac{dV}{dt}
%     =2h(x(t))\pi r_e l_{\text{cap}}(t)(\mathcal{T}_{\text{ext}}-\mathcal{T}(t)).
% \end{aligned}
% \end{equation}
% where $R$, $C_{v}$, $V$, $m_{\text{gas}}$, $\mathcal{T}$, $\mathcal{T}_{\text{ext}}$, and $h(x(t))$ are the specific gas constant, the heat capacity, the gas volume, the gas mass, the gas temperature, the external temperature, and the heat transfer rate, respectively. Note that for different ranges along the tube the external temperature may be set up on different (constant) values such as $\mathcal{T}_{\text{cold}}$, $\mathcal{T}_{\text{hot}}$, and $\mathcal{T}_{\text{atm}}$ for cold, hot, and room temperatures. The heat transfer rate may also differ along the tube according to the governing process corresponding to the concrete region. For further details, see SI, Section 4.

\subsection{Controlling the motion, extension, and compression}
Theoretical simulations according to the model described in Section~\ref{S:motion}, yielded the positions of the solenoids along the tube as well as their actuation type, which allow continuous and programmable motion of the capsules in a closed loop. In Fig.~\ref{Fig3}(A1), (A2), (A3) and (A4), we show a sequence of frames from a video (see Movie S1) in our experiment, where we track the location of the green capsule at times $t=0,\,2.5,\,5.5$, and 8.5 seconds, respectively. As can be seen, the cycle, whose length is 4.34 meters ended after about 8.5 seconds. In this experiment the maximal current in each solenoid was 16 [A] and the number of capsules was 30. The actuation of the solenoids numbers 4, 5, and 6 is the same as of the solenoids numbers 1, 2, and 3, respectively (for the numbers of the solenoids, see Fig.~\ref{Fig3}(A1)). Each actuation cycle, which we will refer hereafter as the ``motion regime,'' consists of four stages with a period of 70 [ms], which yields the frequency of $\omega=1/0.28=3.57$ [Hz]. At the first stage of the ``motion regime,'' the current in solenoids numbers 1 and 3 is 16 and -16 [A], respectively, and solenoid number 2 is off. At the second stage, the current in solenoid number 2 is 16 [A] and solenoids numbers 1 and 3 are off. At the third stage, the current in solenoids numbers 1 and 3 is -16 and 16 [A], respectively, and solenoid number 2 is off. At the fourth stage, the current in solenoid number 2 is -16 [A] and solenoids numbers 1 and 3 are off. Note that our system is very robust, in the sense that even if during the operation regime of motion, one, two, three, or four solenoids are suddenly turned off, still the motion of the metafluid continues with the same velocity (for additional details see Movie S1 and Fig. 5 in Section 4 in SI).   
Moreover, the theoretical actuation followed a similar pattern, as described by equations (8)-(10) in Section 1 in SI.

To validate our model, we compared the theoretical predictions with experimental data. In Fig.~\ref{Fig3}(B), we show a comparison between theory (dashed lines) and experiment (solid lines) for the location of capsules versus time for different frequencies and various numbers of capsules within the tube. All frequencies and numbers of capsules in the system, which were examined, show a linear relation between the location of the capsules and time. In all cases the quantitative agreement between theory and experiment is excellent.
Note that the maximum standard deviation is $\pm 0.11$ [s], and thus the error bars were multiplied by 10, in order to make them visible.  In Fig.~\ref{Fig3}(C), we show the mean velocity of the capsules versus frequency, for different numbers of the capsules within the tube. It can be observed that in all examined cases the velocity is approximately a linearly increasing function of frequency and a decreasing function of the number of capsules in the system. As previously, we obtained an excellent quantitative agreement between theory and experiment. Note that the maximum standard deviation is $\pm 0.015$ [m/s], and thus the error bars were multiplied by 5, in order to make them visible.   

Furthermore, we examined the capsules' expansion and compression regime during the motion in the closed loop. In Fig.~\ref{Fig4}, we show the results of simulation and experiment in the system with 30 capsules. Here, we concentrate on the expansion and compression of the capsules, which occur during their motion in the closed loop and where initially the capsules are in the expanded state. Specifically, in Fig.~\ref{Fig4}(A), (B), and (C), we show the simulation results for initial permutation of the capsule $\overrightarrow{per}=(7,0,11)$ and the currents of $I_{\text{curr}}=2$, 8, and 16 [A], respectively. The actuation functions of solenoids 1, 2, and 3 are shown in Fig.~\ref{Fig4}(E).  Note that in Fig.~\ref{Fig4}(C) the solenoids numbers 4, 5, and 6 were actuated the same as solenoids 1, 2, and 3, respectively, and in panels (A)-(B) with opposite signs relative to solenoids 1, 2, and 3. After one cycle was completed, the capsules' lengths were averaged at each position within the tube and plotted as a height map over the tubes configuration. Panels (A)-(C) visualize the fact that the current intensity in solenoids affects the variation of the average capsules' length. In particular, for a sufficiently small current the variation in the average capsules' length within the tube is negligible. As we increase the current, shown in panels (B)-(C), the average capsules' length becomes less homogeneous and at the current of 16 [A] the difference between the average capsules' length reaches more than 2 [cm], which is more than 25\% of the initial capsules' length. Thus, by programming the  solenoids' actuation, we may control the average capsules length, which may depend on location within the tube and time, according to the demands of a specific application.

To experimentally validate these findings, we performed an additional experiment under similar conditions. In Fig.~\ref{Fig4}(D), we show a frame from our experiment for the expansion and compression of the capsules. The increased view in Fig.~\ref{Fig4}(D1), illustrates the fact that the capsules can achieve different permutations in our experiment. The experiment was performed at the current of $I_{\text{curr}}=16$ [A] and frequency of $\omega=3.57$ [Hz] as follows. Initially, we applied vacuum in the system and thus the capsules expanded to the permutation of $\overrightarrow{per}=(5,0,13)$. Next, the system returned to atmospheric conditions and the capsules reached the stable permutation of $\overrightarrow{per}=(6,0,12)$. Lastly, we actuated the solenoids according to the ``motion regime,'' as described above. We performed a complete cycle and obtained that some capsules compressed to the permutation of $\overrightarrow{per}=(9,0,9)$, where other capsules either remained at the permutation $\overrightarrow{per}=(6,0,12)$ or expanded and/or compressed and then returned back to the permutation $\overrightarrow{per}=(6,0,12)$. 

Overall, our results indicate that both, motion and expansion/compression, regimes may be programmed locally as needed by using external actuation according to the proposed framework.

\section{Concluding remarks}

We demonstrate in this work an approach for employing multi-stable metamaterials in applications involving internal flow. The concept requires the ability to control the movement, expansion, and compression of multiple metamaterial units submerged in a fluid. This type of metafluid, however, cannot be actuated by conventional mechanisms, such as pumps and expansion valves. The presented approach employed time-varying magnetic fields, which are induced by multiple solenoids, as well as magnets attached to the capsules' ends. We demonstrated the ability to move, expand, and compress multi-stable capsules in a programmable manner, thereby producing multi-stable metafluids whose properties can be varied spatially and temporally. Additionally, we developed a theoretical model that showed excellent quantitative agreement with experimental results.

Our approach should be possible to broaden in order to account for additional effects, such as heating or cooling of the encapsulated gas during expansion or compression of the capsules. The rate of heat transfer from the gas to the external environment needs to be studied. Other future directions include incorporation of phase change within the capsules, either liquid-gas or sublimation solid-gas. Optimizing the resulting metafluid will allow our programmable method for manipulating metafluids to incorporate metamaterials in various applications which need fluid-like behavior, such as energy storage and harvesting, cooling cycles, heat engines, and robotic mechanisms.

\section{Materials and methods}

This study aims to utilize time-varying magnetic fields to externally actuate liquefied metamaterials for fluid-like applications. Our device was constructed from a long and closed tube which consist of integrated and sealed Perspex tubes, four 3D-printed corners, and six customized solenoids. The tube is filled with water and multistable capsules, which encapsulate air under atmospheric conditions and are equipped with magnets on both ends. This device is actuated with six power supplies, six drivers, a breadboard, and an Arduino. See Fig.~\ref{Fig2}(A).

\subsection{Fabrication of the multistable capsules and the experimental setup}

The four corners were printed using an SLA 3D printer (Form 3BL). One corner is connected to a fluid reservoir which allows to fill the system with water as needed, while the other three corners are equipped with taps which enable to release air and seal the system. To fabricate the long closed tube, we used Perspex tubes with an inner and outer diameters of 32 and 40 mm, respectively, and of four different lengths of 340, 270, 180, and 170 mm. The custom-designed solenoids were fabricated using a copper tube with an outer diameter of 35 mm and an inner diameter of 31.5 mm as the core. A 1.12 mm diameter enamelled copper wire, with a total mass of 1.5 kg, was wound around the copper tube to form the solenoid. To facilitate integration with the rest of the system, two 3D-printed end components, were printed by using SLA 3D printer (Form 3BL) and adhered to both ends of the solenoid. All solenoids, tubes, and 3D printed corner components were assembled using threaded plumbing connectors shown in Fig.~\ref{Fig2}(A), ensuring a secure and modular integration of the system.

The multistable capsules are composed of 18 segments of a bendy straw with an outer diameter of 29 mm and an inner diameter of 23 mm. Each capsule was sealed with hot glue under atmospheric conditions with 9 segments open and 9 segments closed. Two N52 neodymium magnets, each with a diameter of 25.4 mm and a thickness of 6.35 mm, were attached to both ends of each capsule. Additionally, two hemispherical cups, printed using an SLA 3D printer (Form 3B+), were attached to both sides of the capsule, enclosing the magnets, to reduce friction within the system.

We used six RIDEN RD6024 power supplies, six Cytron MD25HV drivers, a breadboard, and an Arduino MEGA 2560 to control and program the system actuation as desired. The photo in Fig.~\ref{Fig2} and the frames of the experimental setup in Figs.~\ref{Fig3} and~\ref{Fig4} are parts of the videos; all of which were captured using iPhone 15 Pro Max camera in 4K resolution at 60 fps.

\section*{}
{\textbf{Author contributions:}}
A.D.G. and E.B.A. conceived the research subject. E.B.A. constructed the experimental setup and conducted the experiments. A.Z performed the theoretical analysis and numerical computations. E.B.A. analyzed the experimental data. E.B.A., A.Z., S.G., and  A.D.G. wrote the paper.

\bibliography{bibliography}
\bibliographystyle{plain}

\end{document}